\documentclass[11pt]{article}
\usepackage[para,online,flushleft]{threeparttable}
\usepackage{moreverb,url, multicol, soul, graphicx, latexsym, amsmath, amssymb, booktabs}
\usepackage[margin=1.in]{geometry} %to handle margins
\usepackage[
english,
noneedspace
]{SASnRdisplay}
\usepackage[title]{appendix}
\usepackage[colorlinks,bookmarksopen,citecolor=red,urlcolor=red]{hyperref}
\begin{document}
%\runninghead{Atem and Matsouaka}
%\rightline{\today}
\title{Linear regression model with a randomly censored predictor: Estimation procedures}
\author{Folefac D. Atem\affilnum{1} and Roland A. Matsouaka\affilnum{2,}\affilnum{3}}
%\affiliation{\affilnum{1}Department of Biostatistics, University of Texas Health Science Center at Houston, Houston, TX, USA.\\\affilnum{2}Department of Biostatistics and Bioinformatics %, Duke University, Durham, NC, USA. 
%\affilnum{3}Duke Clinical Research Institute, Duke University, Durham, USA.}
%\corrauth{Atem D. Folefac, Department of Biostatistics\\ University of Texas Health Science Center at Houston
%Houston, TX 77030, USA.}
%
%\email{Folefac.Atem@UTSouthwestern.edu}
%\title{A Small \LaTeX{} Article Template\thanks{To your mother}}
\author{Folefac D. Atem \\
	Department of Biostatistics, \\University of Texas Health Science Center, Houston, TX, USA. \\\\
	\and 
	Roland A. Matsouaka \\
	Department of Biostatistics and Bioinformatics\\ \& Duke Clinical Research Institute, Duke University, Durham, USA. \\
	}

\date{\today}
\maketitle
\begin{abstract}
We consider linear regression model estimation where the covariate of interest is randomly censored. Under a non-informative  censoring mechanism, one may obtain valid estimates by deleting censored observations. However, this comes at a cost of lost  information and decreased efficiency, especially under heavy censoring. Other methods for dealing with censored covariates, such as  ignoring censoring or replacing censored observations with a fixed number, often lead to severely biased results and are of limited practicality. Parametric methods based on maximum likelihood estimation as well as semiparametric and non-parametric methods have been successfully used in linear regression estimation with censored covariates where censoring is due to a limit of detection. %Statistical model usually employed either a parametric or non-parametric algorithm. The parametric model makes strong assumption about the data and perform well if the assumptions turns out to be correct. In contrast, the non-parametric algorithm makes fewer assumptions about the data. 
In this paper, we adapt some of these methods to handle randomly censored covariates and compare them under different scenarios to recently-developed semiparametric and nonparametric methods for randomly censored covariates. Specifically, we consider both  dependent and  independent randomly censored mechanisms as well as the impact of using a  non-parametric algorithm on the distribution of the randomly censored covariate. Through extensive simulation studies we compare the performance of these methods under  different scenarios. Finally, we illustrate and compare the methods using  the Framingham Health Study data to assess the association between  low-density lipoprotein (LDL) in offspring and parental age at onset of a clinically-diagnosed cardiovascular event. 

\end{abstract}
%\keywords{Censored covariate; random censoring; maximum likelihood estimation; substitution methods; imputation methods; Complete Case %; Closed test procedure; Intersection-union test; Union-intersection test}

\section{Introduction}
Modeling continuous outcome data using linear regression usually assumes in theory that the values of the covariates are fully observed. However, in practice and especially for any large data set, it is unlikely that complete information will be available for all study participants.  The issue of  censored data is ubiquitous and affects many studies and permeates a wide range of research areas, including medicine, economics, and social sciences . 

Multiple reasons lead to incomplete observations in a data set including nonresponse, attrition, and absence of event of interest. Usually during the design, implementation, and data collection phases of a study, efforts are made to minimize the occurrence of incomplete data whenever possible and, when unavoidable, to understand the reasons for such a discrepancy in order to handle the available data adequately and run appropriate statistical analyses.  %Even in a carefully designed and executed study, it is almost certain that some information will either be missing or in an expandable form. 
Although there is an extensive literature on missing data\cite{little1992regression, little2014statistical, wang2012multiple} and censored outcomes,\cite{lagakos1979general,kleinbaum1996survival,klein2005survival} only a small number of papers have explored scenarios in which the covariate is censored.\cite{cole2009estimating, atem2015linear,rigobon2009bias,rigobon2007estimation} %Methods known to work very well with missing covariate measures such as multiple imputation do not fare well with censored covariates;\cite{lynn2001maximum,arunajadai2012handling} thus the need to developed suitable methods for randomly censored covariates.

Arguably, the inadequacy of  linear regression models on censored outcome  variables has sparked an interest in alternative methods, and subsequently has led to major developments of regression models for survival analysis for decades. Extensive literature has been published regarding censored outcomes, especially in studies of time-to-event outcomes where censoring is due to loss of follow-up, drop out, or study termination.\cite{lagakos1979general, kleinbaum1996survival, klein2005survival}  

While there is a vast literature on censored outcomes and different related methods have been discussed extensively, only a limited number of papers have focused on the issue of censored covariates. Ignoring or using a wrong approach to account for the censored nature of a covariate in  regression model estimation can lead to analytical issues and spurious results \cite{rigobon2009bias, rigobon2007estimation, austin2004estimating}. It is important that censored covariates be recognized, acknowledged,   and handled appropriately to produce reliable results.\cite{helsel2005nondetects,helsel2006fabricating,helsel2005more, lynn2001maximum,arunajadai2012handling,richardson2003effects,albert2010use,hornung1990estimation,sampene2015imputing} However, the vast majority of literature on censored covariates has focused on censoring due  limit of detection or type 1 censored covariates where observations of the covariate below such a limit cannot be measured or detected, but recorded at or less than the limit of detection value.\cite{helsel2005nondetects, bernhardt2014flexible,sattar2015frailty,austin2004estimating, bernhardt2013statistical, wu2012bayesian, lee2012multiple}  Only a handful of publications have investigated the implications of randomly censored covariates where some observations of the covariate are censored at varying censoring time points.\cite{sampene2015imputing,atem2016improving,atem2015linear,atem2016multiple, lee2003proportional}

Censored covariate measurements arise when, for some participants in a study, the ascertained information of interest has not yet occurred (or will not occur) at the time of assessment. This is due to a time lag between the time when a covariate is measured (usually at baseline)  and the occurrence (or non-occurrence) of an event of particular interest that needs to happen for such a measure to be available and assessed. For instance, Clayton \cite{clayton1978model}  investigated familial aggregation in chronic disease incidence and modeled  the possible influence that parental age at onset of a given disease  might have on an individual's risk of succumbing to a particular disease. 
Using the Framingham Heart Study---an ongoing multi-generational landmark  study designed to identify factors and characteristics that contribute to the development of cardiovascular disease (CVD) and other diseases through long-term, active surveillance and monitoring---Atem and Matsouaka \cite{atem2016improving} studied the impact that age at onset of clinically-diagnosed cardiovascular events in parents may have on the  onset of cardiovascular events among offspring. \\
In both cases, even if important factors have been thoroughly measured for both parents and their offspring,  it is unlikely that all parents have had or will have developed the disease of interest at the time of investigation. This means that  the variable "parental age at onset of a given disease" is guaranteed to be censored, i.e., not fully observed. Therefore, it is extremely important in any statistical analysis to account for the fact that the variable of interest is  censored for some participants.

In theory, there are many ways to address the issue of censored covariates in data analyses. From a practical point of view, however, the most important questions are: {\it When and under what conditions can one safely consider the problem of censored covariates to be trivial? How can current methods be applied and under what conditions can one expect (asymptotically) unbiased and meaningful results?}
In general, inappropriate handling of censored covariates may affect the type I error\cite{austin2004estimating}, yield biased results,  hinder the power to detect any meaningful treatment differences, or lead to loss of efficiency  in estimating the coefficient parameters of a regression model.\cite{rigobon2007estimation} 

 Complete-case analysis, whereby observations with censored covariate values are discarded (on purpose or through a software default option), is the most commonly used method. When the sample size of the data is large, the censoring mechanism is independent of the outcome, and the proportion of censored data is relatively small, complete-case analysis of the data can be employed safely\cite{little1992regression, little2014statistical} since the impact of censored observations on the analysis of the data may be negligible.  In that case the complete-case analysis yields valid (consistent and asymptotically unbiased) estimates for regression coefficient parameters. 
 
 However, under moderate to heavy censoring of data there might be a substantial loss of efficiency due to the reduction in the sample size and the significant loss of information on other fully-observed covariates and on the outcome measures of the deleted observations.\cite{lipsitz2004generalized}  Furthermore, when the censoring mechanism is informative, using a complete-case analysis can lead to biased results which are in part exacerbated by the losses of information and efficiency since restricting the analysis solely to truly observed covariate measures may introduce some  imbalance in the dataset in a way that misrepresents the  population under study.

When the issue of censored covariate is not ignored all together, simple substitution methods (or to be more accurate {\it ad hoc} fill-in methods)---where censored observations are replaced by the overall mean or median of the observed  variable or, alternatively, by a constant---are  frequently used because  they are simple, easy to understand,  are easy to implement. Unfortunately, they may lead to substantial biased estimates and inaccurate conclusions.\cite{d2008index, bernhardt2014flexible,sattar2012parametric, sattar2015frailty, may2011maximum,rigobon2009bias,helsel2006fabricating} 

 Several non-trivial statistical methods have been developed specifically to input  censored covariates and estimate regression coefficient parameters in a model with a censored covariate.\cite{vexler2015toolkit, kong2016semiparametric} Some of these methods, known as  parametric methods, use maximum likelihood estimation (MLE) under the assumption that the covariate follows a specific distribution.\cite{nie2010linear, arunajadai2012handling, langohr2004parametric, sattar2012parametric, lee2003proportional, d2008index, bernhardt2014flexible, cole2009estimating, albert2010use, schisterman2006limitations, schisterman2010opening,kong2016semiparametric} For example, when such a distribution assumption is plausible, Richardson and Ciampi\cite{richardson2003effects} proposed using MLE and input censored observations with $E(X|X<\xi),$ in the context where measurements of the covariate,  $X,$ are left-censored  due to limit of detection $\xi.$ However, this approach has some limitations, especially when the censored covariate is correlated with other covariates.\cite{nie2010linear, kong2016semiparametric, schisterman2006limitations, schisterman2010opening} %Schisterman et al. \cite{schisterman2006limitations} suggest replacing censored observations with $E(X|X\geq \xi)$ and noticed that neither distributional nor asymptotic assumptions are required to estimate this quantity. 

As we know, an  MLE method relies on a parametric distribution assumption of the censored covariate, i.e., the postulated distribution is assumed to be true and correctly specified. It is less reliable when the distribution assumption is incorrect or when the data set is so small that it becomes questionable whether the assumed distribution fits the data well. In that case, a semiparametric model that makes weaker parametric distribution assumptions  or, even better, a nonparametric method that does not assume any specific distribution model at all is preferable.\cite{nie2010linear,kong2016semiparametric,lynn2001maximum}  %This is the approach adopted by Atem et al. \cite{atem2015linear, sampene2015imputing, atem2016multiple, atem2016improving}

% In this paper,  we looked at two types of censored covariate: (1) dependent censoring, i.e. when the probability of censoring depends on the outcome and (2)  independent censoring, i.e. when the probability of censoring does not depend on the outcome.  

As we previously mentioned, most of these methods highlighted above have been developed to account for type I  censoring or limit of detection and are typically developed for left-censored covariates. Nevertheless, it is fairly straightforward to adapt the methods for a  limit of detection  data or  type I censored covariate to a right-censored covariate. However, to the best of our knowledge, no such parametric approach employed for type I censored covariates  has been extended to handle a randomly censored covariates. In addition, barring a few papers on dependent (randomly) censoring mechanism\cite{sampene2015imputing}, the vast majority of published methods for type I  censoring rely on the assumption that the censoring mechanism is independent of the outcome in interest.\cite{hornung1990estimation, lynn2001maximum, may2011maximum, helsel2005nondetects} 

Our primary objective in this paper is to adapt a parametric method proposed in linear regression models with a type I  censored covariate to the case in which a covariate is randomly censored covariate. We use simulation studies to compare this newly developed method to the methods proposed by Sampene and Folefac \cite{sampene2015imputing}---in which randomly censored covariate values are replaced by a nonparametric and a semiparametric estimations of $E(X|X\geq \tau)$ or $E(X|X\geq \tau, Y)$, where $\tau$ denotes the maximum observation time for the variable $X$, and the outcome of interest $Y$.
For this purpose, we will  consider %for a covariate subject to % either independent or dependent censoring. We will consider 
both dependent and independent censoring mechanism which occurred depending on whether such a censoring mechanism depends or not on the outcome of interest. %(i.e., where, conditional on the vector of fully measured covariates, the censoring mechanism does not depend on the outcome of interest) and dependent censoring,  where such a mechanism depends on the outcome. 
Furthermore, we will also compare the aforementioned  nonparametric  estimation method to  the commonly used deletion or complete-case analysis  and give recommendations on the methods of estimation based on our simulation results.

%Keeping in mind, however, the varied needs of researchers, we will organize our discussion around the following practical issues: {\color{red} Re-write this sentence} (1) while the simple deletion, substitution, and mean imputation are not appropriate for randomly censored covariate, and (2) when the parametric and non-parametric algorithms might be good for randomly censored covariate. {\color{red} Do you mean \lq\lq{(1) when the simple deletion, substitution, and mean imputation are not appropriate for randomly censored covariate, and (2) when the parametric and non-parametric algorithms might be good for randomly censored covariate}?\rq\rq{}} or do you mean  {\color{red} \lq\lq{}while the simple deletion, substitution, and mean imputation are not appropriate for randomly censored covariate,  when do the parametric and non-parametric algorithms might be good for randomly censored covariate?\rq\rq{}}

%\subsection{Assessing the nature of a censored predictor}
 We begin in Section \ref{sec:methods} by presenting parametric and non-parametric methods used in the censored covariate literature. We then introduce the methods proposed by Sampene and Folefac \cite{sampene2015imputing}  to handle randomly censored covariates. In Section  \ref{sec:simul} we run simulation studies to compare each of the discussed methods as well as the complete-case analysis method. Finally, we apply these methods in Section \ref{sec: apply} to the Framingham Offspring Study to assess the influence of parental age at onset of cardiovascular disease on the systolic blood pressure of their offspring.
\section{Notation and Methods}\label{sec:methods}
We consider $n$ study participants independently sampled from a referenced population. Let $Y_i$, $X_i$ and $C_i$ be, respectively, the continuous outcome variable, the potentially censored covariate (from which we are interested in making inferences), and  the right censoring variable, where $i$ indexes subjects. 

Due to the right-censoring in the covariate $X$, for each participant $i$, we observe the vector $(Y_i, V_i,\delta_i)^\top$ where $V_{i}=\min(X_{i}, C_{i})$,  $i=1, 2, \ldots, n$. The linear regression model is given by
 \begin{equation}\label{eq:main_eq}
  Y=\beta_0+ \beta_1 X +\epsilon    
  \end{equation}
 where  the parameter coefficients $\beta_0$  and $\beta_1$   are the intercept and slope, respectively. The random error $\epsilon$ is assumed to be independent of $X$ and follows a normal distribution with mean 0 and variance $\sigma^2$  i.e. $\epsilon$ $\sim N({0},\sigma^2)$.  
 
We consider two different cases of censoring mechanisms. In the first case, we assume that the censoring mechanism is non-informative, i.e., $C$ is independent of the outcome $Y$. For the second case, we assume that the censoring mechanism depends on the outcome $Y,$ in a sense that there is some known point $c_0 \in \mathbb{R}$ such that the random variable $C$ follows a distribution characterized by the distribution function $G_0$
when $Y < c_0 $ and by the distribution function $G_1$ when $Y\geq c_0.$

%We assumed $X$ may be subject to random censoring, i.e., $X_i=min(T_i,C_i)$.  Censoring is  independent if $C_i$  is independent of $Y_i$ that is $C=C_i$ for all $y_0\in Y$   and it is dependent if $C_i$  depends on  $Y_i$  i.e.: 
%\begin{equation}
%c = \left\{
%\begin{array}{lr}
%G_0 & : y < y_0\\
%G_1 & : y \ge y_0
%\end{array}
%\right.
%\end{equation}
%where  $y_0\in Y$ and $c\in C$ with distribution $G_0$ and $G_1$ based on threshold $y_0$. 
 
%We assumed $X$ may be subject to random censoring, i.e., $X_i=min(T_i,C_i)$.  Censoring is  independent if $C_i$  is independent of $Y_i$ that is $C=C_i$ for all $y_0\in Y$   and it is dependent if $C_i$  depends on  $Y_i$  i.e.: 
% \begin{equation}
% c = \left\{
% \begin{array}{lr}
% G_0 & : y < y_0\\
% G_1 & : y \ge y_0
% \end{array}
% \right.
% \end{equation}
% where  $y_0\in Y$ and $c\in C$ with distribution $G_0$ and $G_1$ based on threshold $y_0$. 
For simplicity and demonstrative purposes, we limit our discussion to cases with no  additional covariates. If, in practice, the data at hand contain a set of additional fully-observed (i.e. non-censored) covariates, $\mathbf{Z}$,  the method discussed here could easily be extended to accommodate such covariates.  %Our goal is to compare parametric and non-parametric methods under both independent and dependent censoring for the variable $X$ that we have outlined in the previous section.
%In this paper we shall compared parametric and non-parametric approaches when $C$ is independent of $Y$ i.e. independent censoring and when $C$ depends on $Y$  i.e. dependent censoring.

\subsection{Parametric method: Maximum likelihood estimation}
%In this section, we present parametric approach for a randomly censored covariate and discuss when they might not be appropriate for a randomly censored covariate. Possible extension to dependent censored covariate is presented provided the distribution of censored observation are known.
%Parametric solution for a left independent censored covariate subject to limit of detection or fixed censored covariate were developed by Lynn 2001\cite{lynn2001maximum} and Nie et al 2011 \cite{nie2010linear}, assuming $X$ to be normally distributed. 

%Supposed the data consist of  $n$ copies $\{(X_i, Y_i)$, $i=1,\ldots, n\}$, sampled from the population of interest. Each $X_{i}$ is subject to random right censoring and $X_{i}$  be i.i.d $\sim \mathcal{N} ({\mu_{x}},\sigma_{x}^2)$.
A parametric method assumes an underlying distribution of the population from which the data at hand were sampled and uses the maximum likelihood estimation method to draw inference. 

Suppose that the censoring $C$ is independent of $Y$; this implies $X_{i}$ and $ \epsilon_{i}$ are independent. Therefore, the distribution of $Y$ is a product of the distributions of $X_{i}$ and $\epsilon_{i}$.
The likelihood $L$ of $Y$ is made up of two components; one based on the uncensored (observed)    $X_{1},\ldots, X_{m}$ and the other on the right-censored
 $X_{m+1},\ldots, X_{n}$: 
\begin{equation*} L = \prod\limits_{i=1}^{m}f(Y_{i},X_{i})^{\delta}\prod\limits_{i=m+1}^{n}f(Y_{i}, X_{i}>c_{i})^{1-\delta},
\end{equation*}
with $\delta=1$ if $X_i$ is observed and $\delta=0$ if $X_i$ is censored, $i=1,\ldots, n.$ The maximum likelihood estimate of the unknown regression parameter corresponding to the censored covariate $X$ is derived from the log-likelihood function
\begin{equation} \label{eq:loglik}
\log(L) = \delta\sum\limits_{i=1}^{m}\log(l_{i1}) + (1-\delta)\sum\limits_{i=m+1}^{n}\log(l_{i2}),
\end{equation}
where $\log(l_{i1})=\log[f(Y_{i},X_{i})]$ and $\log(l_{i2})=\log[f(Y_{i}, X_{i}>c_{i})].$

Suppose that $X$ follows a normal distribution with mean $\mu_{x}^2$ and variance $\sigma_{x}^2,$ we have %the equation \eqref{eq:loglik} becomes a sum of two components (the unconsored and censored components) we will define shortly. The 
 %observed component$ l_{i1}$ is thus equal to 
\begin{equation*} l_{i1}=\frac{1}{2\pi \sigma \sigma_{x}} \exp\bigg({ \frac{-\epsilon_{i}^2}{2\sigma^2}-\frac{(x_{i}-\mu_{x})^2}{2\sigma_{x}^2}}\bigg).\end{equation*}
For the censored component,  consider $\Phi(x)$  the cummulative gaussian distribution function and define
 $\epsilon_{i}=y_{i}-\beta_{0}-\beta x_{i}$,  $\epsilon_{i\mu_x}=y_i-\beta_{0}-\beta\mu_{x},$ and  $Q=\sqrt{\sigma_{x}^{-2}+\beta^2\sigma^{-2}}.$ We show in the Appendix that 
\begin{eqnarray}
f(Y_{i},X_{i}> c_{i})
=
\frac{1}{\sqrt{2\pi(\sigma^2 + \beta_1^2 \sigma_x^2)}}
\Phi
\left\{
Q \left(
 \mu_x + \frac{\beta_1\epsilon_{i\mu_x}}{\sigma^2 Q^2}-c_i 
\right) \right\}\exp\left\{ - \frac{\epsilon_{i\mu_x}^2}{2\sigma^2}
\left( 1 - \frac{\beta_1^2}{\sigma^2 Q^2} \right) \right\}.
\end{eqnarray}

When censoring is dependent as described above, $C_{i}$ and $Y$ are dependent. The likelihood can be expressed as
\begin{equation} L = \prod\limits_{i=1}^{m}f(Y_{i},X_{i})\prod\limits_{i=m+1}^{k}G_0(Y_{i}, X_{i}>c_{i})\prod\limits_{i=k+1}^{n}G_1(Y_{i}, X_{i}>c_{i}),
\end{equation}
where the data is made of  fully observed   $x_{1},\ldots, x_{m}$  and right censored $x_{m+1},\ldots, x_{n}$. The censored component is divided into $x_{m+1},\ldots, x_{k}$ and $x_{k+1},\ldots, x_{n}$ components with associated distribution $G_0$ and $G_1$ respectively, as in equation (2). %If $G_0$ and $G_1$ are normally distributed as most likely assumed, the likelihood reduces to equation (4) above.

\subsection{Overview of nonparametric methods}
As stated in the introduction, most of the methods described  in published literature that examine the issue of covariates  are subject to the limit of detection. Prior to the late 1990's, the most common approach to handling such censored covariates was the complete-case analysis method.% in which censored observations were simply deleted. When the censoring mechanism is non-informative, this approach results in unbiased estimates, but with a loss in efficiency\cite{little2014statistical}. 

Alternatively, several naive {\it ad hoc} alternative methods have been proposed, including substitution methods, which consist of replacing  censored covariate values with either a function of the limit of detection, $\xi$, e.g.,  $\xi$, $\sqrt{2}\xi$, ${2}\xi$ \cite{hornung1990estimation} or the  mean of the observed covariate measures (mean substitution)\cite{schisterman2006limitations} %\cite{schisterman2006limitations, tsimikas2012inference} for fixed censored design in covariates subject to limit of detection. 
%In the latter approach censored values are simply replaced  by the .  %These approaches are most popular for censored data subject to limit of detection.  
as well as dichotomizing the potentially censored covariate into a binary covariate.\cite{austin2004estimating,rigobon2009bias} Inevitably, each of these {\it ad hoc} methods leads to a biased estimation of $\beta_1$. 
For instance, Helsel  investigated the use of these naive substitution methods and concluded that they are inefficient and  have no mathematically plausible backing \cite{helsel2005more}.  The extent of the inefficiency depends on the extent and the severity of censoring (i.e., the distance between the limit of detection or random censoring value and the natural limit for $X$) of censoring.  % This approach is not widely used since it was shown to be heavily bias \cite{austin2004estimating, rigobon2009bias}. 
Finally,  Atem et al\cite{atem2015linear, atem2016multiple} explored additional non-parametric methods, based on  multiple imputation approach, but  concluded that these methods were not efficient when applied to the cases of dependent censoring. % {\color{red} Are these non-parametric MI methods based on Rubin's approach? If yes, then we don't need to mention Rubin's approach at the end of the next paragraph, but move it and integrate it to this sentence. If not, we should remove this sentence from here as it doesn't help to talk about an approach we won't use and which is not common like Rubin's. Or even better, we should present the algorithm of these nonparametric MI here and give the related references, which will give a sense to the sentence written at the end of the next paragraph}

Recently, Sampene and Atem \cite{sampene2015imputing} proposed two conditional multiple imputation methods for estimation and inference. The underpinnings of these methods involve  replacing the randomly censored values $X_i$ by estimates of $E(X_i|X_i>\tau)$, for $i=m+1\ldots, n$. In the absence of additional covariates, the former is determined via a Kaplan-Meier estimator and performs well  when the correlation between $X$ and $Y$ is weak.  When the correlation  between $X$ and $Y$ is strong, similar to case of missing covariate \cite{little1992regression} the outcome of interest is included in the  imputation  $E(X_i|X_i>\tau, Y_i)$. This conditional imputation involving outcome $Y$, unlike the imputation not involving $Y$ used estimates from the Cox proportional hazard hence the name Cox Multiple Imputation. To estimate the corresponding variance of $\beta_1$ for inference, Sampene and Atem \cite{sampene2015imputing} suggested using either a conditional multiple imputation or a conditional single imputation along with a bootstrap resampling procedure to correct for the underestimation of the variance inherent to the single imputation. In doing so, we showed that these improvements to the complete-case analysis method result in valid inferences regardless of whether the censoring mechanism is dependent or independent of the outcome. Furthermore, using simulation studies, they demonstrated that the multiple imputation method is similar to the conditional single imputation with bootstrap resampling.  % These two distributional free approaches are different from Rubin's \cite{rubin2004multiple} multiple impuation approach in that, it does not involve simulated data in solving the numerical solution. 

In the next section, we will run simulation studies to compare the complete-case analysis, parametric, mean imputation, naive {\it ad hoc}, conditional single imputation,  conditional multiple imputation and Cox multiple imputation methods. It is worth mentioning that Cole et al (2009) and Nie et al  \cite{nie2010linear} have explored the parametric approach for type 1 censoring. They showed that this approach is very efficient for limit of detection data. However, as pointed out by one of the reviewers, it is worth exploring how well this parametric approach works compared to others non parametric approaches when censoring is random.  
%\section{ Results}

\section{Monte Carlo Simulations}\label{sec:simul}
\subsection{Data generation and simulation set up}
%To compare the methods highlighted in the previous section, we conducted extensive simulation studies. 
We assumed that the true linear regression model is given by $Y=\beta_0+ \beta_1 X +\epsilon,$    with  $(\beta_0, \beta_1)$ = $(1,0.5)$ and $\epsilon\sim N(0,1) $.  The variable $X$ as well as the censoring variable distribution were generated from a two-parameter Weibull distribution
\begin{equation}
f(X)=\frac{\theta}{\lambda}\bigg(\frac{X}{\lambda}\bigg)^{\theta-1}\exp\left[-\bigg(\frac{X}{\lambda}\bigg)^\theta\right]
\end{equation}
where
$\theta$ is the shape parameter also known as the Weibull slope, with $\theta>0$ and 
$\lambda$, $\lambda>0$, is the scale parameter.
%This Weibull distribution was used because of the special properties of the distribution of the Weibull as the shape parameter $\theta$ increases. Figure 1 below shows the distribution of the Weibull for: $0<\theta<1$, $\theta=1$ and $\theta>1$. The censored distribution was also simulated from the same distribution with different parameters.

More precisely,  we generated $K=2000$ samples of size $n=100$ and $n=500$ respectively, and chose $\epsilon\sim N(0,1)$ in each case. For independent censoring mechanism, we considered the following distributions
\begin{itemize}
	\item $X \sim Weibull(3/4,1/4)$ and $C\sim Weibull(1, q)$ for $q=1.50$ and $0.35$. %We generated 2000 samples from our data generating process.
	\item $X \sim Weibull(1,1/4)$ and $C\sim Weibull(5/4, q)$ with $q=1$ and $0.40$. % We generated 2000 samples from our data generating process.
	\item $X \sim Weibull(2,1/4)$ and $C\sim Weibull(9/4, q)$,  $q=0.50$ and $0.30$. %We generated 2000 samples from our data generating process.
\end{itemize}
The selected values of $q$  allowed us to obtain, respectively, $20\%$ and $40\%$ censoring.
Under dependent censoring, we defined  the corresponding mechanism such that  $C=C_1$ if $Y>1.02$ and $C=C_2$ if $Y\le 1.02$. We also considered the following data generating distributions  to obtain $20\%$ and $40\%$ censoring: $X \sim Weibull(2,1/4)$, $C_1\sim Weibull(9/4, q)$ and $C_2\sim Weibull(10/4, q)$ with $q=0.50$ and $0.30$, respectively. 
 
 \subsection{Simulation results}
 Tables \ref{tab:1Ba}--\ref{tab:1Bd} summarize the results of the four sets of simulations performed for light censoring (20\%) and  heavy censoring (40\%) in terms of
 \begin{enumerate}  \item $\displaystyle Bias=\left(\frac{1}{K}\sum_{k=1}^{K}\widehat{\beta}_{1k}\right)-\beta_1$, which is the (overall) deviation of a parameter estimate from the true parameter $\beta_1=0.5,$   where $\widehat{\beta}_{1k}$ is the estimate from the $k$-th generated data set;  
 \item empirical standard error, $SE(\widehat{\beta}_1)$, of the estimate $\widehat{\beta}_1$ over all $K$ simulation data sets; 
 \item simulation error, i.e.,  the average of model-based standard errors; 
 \item mean squared error (MSE), which is the expectation of the square deviation of a parameter estimate from the truth. It is equal to  $Bias^2+SE(\widehat{\beta}_1)^2$;
 \item coverage probability which is the proportion of simulated samples for which  the $95\%$ confidence interval $\widehat{\beta}_{1k}\pm Z_{1-\alpha/2}SE(\widehat{\beta}_{1k})$ includes $\beta_1=0.5,$ for $k=1,...,K$.
 \end{enumerate}
 \begin{table}
 	\caption{Case 1:  $X \sim Weibull(3/4,1/4)$ and censoring is independent of  $Y$ .}\label{tab:1Ba}
 	\begin{center}
 		\begin{tabular}{lcl*{5}{c}}
 		\toprule
 			\textbf{Light Censoring} &  &  &  &  & \\ \midrule 
 				& &  & Simulation & & Coverage\\
 			$N=100$ & Bias & $SE(\hat{\beta})$ &  Error & MSE & Probability\\ \cmidrule(lr){1-6}
 						Actual data (No Censoring) & 0.0012 & 0.2527 & 0.2536 & 0.0639 & 0.970\\
 						Complete-case & 0.0044 & 0.4138 & 0.4163 & 0.1712 & 0.955\\
 					%	Ad hoc Substitution & 0.0229 & 0.3163 & 0.3087 & 0.1006 & 0.955\\
 						Mean Substitution & 0.0044 & 0.4200 & 0.4163 & 0.1712 & 0.962\\
 						Maximum Likelihood & 0.0510 & 0.3345 & 0.3377 & 0.1145 & 0.945\\
 						Conditional Single Imputation & 0.0361 & 0.2901 & 0.3138 & 0.0855 & 0.969\\
 						Conditional Multiple Imputation & 0.0014 & 0.4011 & 0.4111 & 0.0842 & 0.960\\
 						Cox Based Multiple Imputation & 0.0013 & 0.4201 & 0.4211 & 0.1765 & 0.966  \\\midrule\addlinespace
 			$N=500$ &  &  &  &  & \\ \cmidrule(lr){1-6}
 						Actual data (No Censoring) & 0.0009 & 0.1103 & 0.1118 & 0.0122 & 0.971\\
 						Complete-case & 0.0021 & 0.1795 & 0.1799 & 0.0322 & 0.955\\
 					%	Ad hoc Substitution  & 0.0113 & 0.1367 & 0.1407 & 0.0188 & 0.940\\
 						Mean Substitution & 0.0021 & 0.1820 & 0.1799 & 0.0331 & 0.957\\
 						Maximum Likelihood & 0.0361 & 0.1469 & 0.1530 & 0.0229 & 0.945\\
 						Conditional Single Imputation & 0.0132 & 0.1297 & 0.1524 & 0.0170 & 0.970\\
 						Conditional Multiple Imputation & 0.0011 & 0.1811 & 0.1815 & 0.0324 & 0.960\\
 						Cox Based Multiple Imputation & 0.0012 & 0.1705 & 0.1743 & 0.0290 &0.966  \\\midrule \addlinespace%\bottomrule
 			\textbf{Heavy Censoring} &  &  &  &  & \\ \midrule\addlinespace
 			$N=100$ &  &  &  &  & \\ \cmidrule(lr){1-6}
 							Actual data (No Censoring) & 0.0012 & 0.2527 & 0.2536 & 0.0639 & 0.970\\
 							Complete-case & -0.0048 & 0.8683 & 0.9052 & 0.7540 & 0.943\\
 					%		Ad hoc Substitution  & 0.0713 & 0.4800 & 0.4891 & 0.2355 & 0.949\\
 							Mean Substitution & -0.0048 & 0.8852 & 0.9052 & 0.7836 & 0.946\\
 							Maximum Likelihood & 0.2007 & 0.5229 & 0.5421 & 0.3137 & 0.900\\
 							Conditional Single Imputation & 0.1000 & 0.4542 & 0.5657 & 0.2163 & 0.930\\
 							Conditional Multiple Imputation & 0.0545 & 0.8101 & 0.7887 & 0.6592 & 0.970\\
 							Cox Based Multiple Imputation & 0.0029 & 0.8700 & 0.8911 & 0.7569 & 0.965 \\\midrule\addlinespace
 			$N=500$ &  &  &  &  & \\ \cmidrule(lr){1-6}
 							Actual data (No Censoring) & 0.0009 & 0.1103 & 0.1118 & 0.0122 & 0.971\\
 							Complete-case & -0.0125 & 0.3744 & 0.3845 & 0.1403 & 0.948\\
 					%		Ad hoc Substitution  & 0.0574 & 0.2116 & 0.2109 & 0.0481 & 0.942\\
 							Mean Substitution & -0.0126 & 0.3820 & 0.3845 & 0.1461 & 0.952\\
 							Maximum Likelihood & 0.0949 & 0.2334 & 0.2369 & 0.6348 & 0.911\\
 							Conditional Single Imputation & 0.0613 & 0.1812 & 0.2188 & 0.0366 & 0.950\\
 							Conditional Multiple Imputation & 0.0060 & 0.3891 & 0.3691 & 0.1475 & 0.961\\  
 							Cox Based Multiple Imputation & 0.0021 & 0.3786 & 0.3888 & 0.1433 & 0.966\\ \bottomrule
 						
 		\end{tabular}
 	\end{center}
 \end{table}%
 
 \begin{table}
 		\caption{Case 2: $X \sim Weibull(1,1/4)$ and censoring is independent of  $Y$ .}\label{tab:1Bb}
 	\begin{center}
 		\begin{tabular}{lcl*{5}{c}}
 	%	\multicolumn{6}{c}{Independent Censoring}  \\ 
 		\toprule
 			\textbf{Light Censoring} &  &  &  &  & \\ \midrule 
 				& &  & Simulation & & Coverage\\
 			$N=100$ & Bias & $SE(\hat{\beta})$ &  Error & MSE & Probability\\ \cmidrule(lr){1-6}
 						Actual data (No Censoring) & 0.0086 & 0.3934 & 0.3969 & 0.1548 & 0.968\\
 						Complete-case & 0.0232 & 0.5419 & 0.5717 & 0.2942 & 0.942\\
 					%	Ad hoc Substitution & 0.0091 & 0.4367 & 0.4600 & 0.1908 & 0.936\\
 						Mean Substitution & 0.0232 & 0.5446 & 0.5717 & 0.2971 & 0.942\\
 						Maximum Likelihood & 0.0510 & 0.4605 & 0.4833 & 0.2147 & 0.939\\
 						Conditional Single Imputation & 0.0185 & 0.4399 & 0.4641 & 0.1939 & 0.966\\
 						Conditional Multiple Imputation & 0.0170 & 0.5818 & 0.5956 & 0.3388 & 0.960\\
 						Cox Based Multiple Imputation & 0.0131 & 0.5463 & 0.5554 & 0.2986 & 0.962 \\\midrule\addlinespace
 			$N=500$ &  &  &  &  & \\ \cmidrule(lr){1-6}
 						Actual data (No Censoring) & 0.0048 & 0.1757 & 0.1731 & 0.0309 & 0.973\\
 						Complete-case & 0.0073 & 0.2426 & 0.2315 & 0.0589 & 0.958\\
 				%		Ad hoc Substitution  & 0.0099 & 0.1942 & 0.1907 & 0.0378 & 0.957\\
 						Mean Substitution & 0.0073 & 0.2436 & 0.2314 & 0.0594 & 0.959\\
 						Maximum Likelihood & 0.0338 & 0.2075 & 0.2047 & 0.0442 & 0.944\\
 						 Conditional Single Imputation & 0.0058 & 0.1927 & 0.2801 & 0.0372 & 0.972\\
 						 Conditional Multiple Imputation & 0.0092 & 0.2481 & 0.2566 & 0.0616 & 0.961\\
 						 Cox Based Multiple Imputation & 0.0060 & 0.2450 & 0.2456 & 0.6001 & 0.962\\\midrule \addlinespace%\bottomrule
 			\textbf{Heavy Censoring} &  &  &  &  & \\ \midrule\addlinespace
 			$N=100$ &  &  &  &  & \\ \cmidrule(lr){1-6}
 			Actual data (No Censoring) & 0.0086 & 0.3934 & 0.3969 & 0.1548 & 0.968\\
 			Complete-case & 0.0334 & 0.8828 & 0.9124 & 0.7805 & 0.941\\
 		%	Ad hoc Substitution  & -0.0101 & 0.5445 & 0.5296 & 0.2966 & 0.957\\
 			Mean Substitution & 0.0334 & 0.8890 & 0.9124 & 0.7914 & 0.941\\
 			Maximum Likelihood & 0.0914 & 0.5967 & 0.6065 & 0.3644 & 0.892\\
 			Conditional Single Imputation & 0.0663 & 0.5466 & 0.6072 & 0.3032 & 0.960\\
 			Conditional Multiple Imputation & 0.0311 & 0.8600 & 0.8340 & 0.7406 & 0.961\\
 			Cox Based Multiple Imputation & 0.0200 & 0.9001 & 0.9004 & 0.8106 & 0.959 \\ \midrule\addlinespace
 			$N=500$ &  &  &  &  & \\ \cmidrule(lr){1-6}
 			Actual data (No Censoring) & 0.0048 & 0.1757 & 0.1731 & 0.0309 & 0.973\\
 			Complete-case & 0.0108 & 0.3893 & 0.3865 & 0.1517 & 0.956\\
 		%	Ad hoc Substitution  & 0.0091 & 0.2438 & 0.2438 & 0.0595 & 0.944\\
 			Mean Substitution & 0.0108 & 0.3922 & 0.3864 & 0.1539 & 0.956\\
 			Maximum Likelihood & 0.0900 & 0.2704 & 0.2767 & 0.0812 & 0.935\\
 			Conditional Single Imputation & 0.0367 & 0.2319 & 0.2564 & 0.0551 & 0.960\\
 			Conditional Multiple Imputation & 0.0121 & 0.4000 & 0.3811 & 0.1601 & 0.961\\
 			Cox Based Multiple Imputation & 0.0061 & 0.3869 & 0.3905 & 0.1497 & 0.965\\\bottomrule %\midrule			
 		\end{tabular}
 	\end{center}
 \end{table}
 
 \begin{table}
 		\caption{Case 3: $X \sim Weibull(2,1/4)$ and censoring is independent of  $Y$ .}\label{tab:1Bc}
 	\begin{center}
 		\begin{tabular}{lcl*{5}{c}}
 %	\multicolumn{6}{c}{Independent Censoring}  \\
 	 \toprule
 			\textbf{Light Censoring} &  &  &  &  & \\ \midrule 
 				& &  & Simulation & & Coverage\\
 			$N=100$ & Bias & $SE(\hat{\beta})$ &  Error & MSE & Probability\\ \cmidrule(lr){1-6}
 			Actual data (No Censoring) & 0.0264 & 0.8322 & 0.8258 & 0.6932 & 0.958\\
 			Complete-case & 0.0465 & 1.0370 & 1.0286 & 1.0775 & 0.948\\
 		%	Ad hoc Substitution  & -0.0436 & 0.8432 & 0.8923 & 0.7131 & 0.954\\
 			Mean Substitution & 0.0465 & 1.0366 & 1.0286 & 1.0767 & 0.948\\
 			Maximum Likelihood & 0.0294 & 0.8894 & 0.8827 & 0.7919 & 0.952\\
 			Conditional Single Imputation & 0.0361 & 0.8855 & 1.4533 & 0.7854 & 0.943\\
 		Conditional Multiple Imputation & 0.0451 & 1.0349 & 1.0381 & 1.0731 & 0.959\\
 		Cox Based Multiple Imputation & 0.0430 & 1.0391 & 1.0396 & 1.0816 & 0.959\\ \midrule\addlinespace
 			$N=500$ &  &  &  &  & \\ \cmidrule(lr){1-6}
 			Actual data (No Censoring) & 0.0008 & 0.3795 & 0.3765 & 0.1440 & 0.970\\
 			Complete-case & 0.0064 & 0.4698 & 0.4703 & 0.2208 & 0.949\\
 		%	Ad hoc Substitution  &- 0.0464 & 0.3898 & 0.3791 & 0.1541 & 0.947\\
 			Mean Substitution & 0.0064 & 0.4698 & 0.4703 & 0.2207 & 0.948\\
 			Maximum Likelihood & 0.0085 & 0.4085 & 0.4132 & 0.1604 & 0.970\\
 			Conditional Single Imputation & 0.0044 & 0.4005 & 0.6782 & 0.1604 & 0.970\\
 			Conditional Multiple Imputation & 0.0019 & 0.5101 & 0.5139 & 0.2602 & 0.961\\
 			Cox Based Multiple Imputation & 0.0021 & 0.4704 & 0.4777 & 0.2212 & 0.969 \\\midrule % \bottomrule
 			\textbf{Heavy Censoring} &  &  &  &  & \\ \midrule\addlinespace
 			$N=100$ &  &  &  &  & \\ \cmidrule(lr){1-6}
 			Actual data (No Censoring) & 0.0264 & 0.8322 & 0.8258 & 0.6932 & 0.958\\
 			Complete-case & 0.0516 & 1.4517 & 1.4531 & 2.1074 & 0.943\\
 		%	Ad hoc Substitution $\sqrt2C$ & -0.0886 & 0.8698 & 0.8997 & 0.7644 & 0.954\\
 			Mean Substitution & 0.0516 & 1.4555 & 1.4531 & 2.1142 & 0.945\\
 			Maximum Likelihood & 0.0204 & 1.0124 & 1.0700 & 1.0254 & 0.941\\
 			Conditional Single Imputation & 0.0371 & 1.0079 & 1.3468 & 1.0172 & 0.942\\
 			Conditional Multiple Imputation & 0.0429 & 1.5321 & 1.5386 & 2.3492 & 0.959\\
 			Cox Based Multiple Imputation & 0.0916 & 1.4573 & 1.4857 & 2.2157 & 0.959  \\\midrule\addlinespace
 			$N=500$ &  &  &  &  & \\  \cmidrule(lr){1-6}
 			Actual data (No Censoring) & 0.0008 & 0.3795 & 0.3765 & 0.1440 & 0.970\\
 			Complete-case & 0.0067 & 0.6537 & 0.6514 & 0.4274 & 0.946\\
 		%	Ad hoc Substitution $\sqrt2C$ & 0.0874 & 0.3952 & 1.0247 & 0.1638 & 0.947\\
 			Mean Substitution & 0.0067 & 0.6538 & 0.6515 & 0.4275 & 0.947\\
 			Maximum Likelihood & 0.0191 & 0.4639 & 0.4585 & 0.2156 & 0.951\\
 			Conditional Single Imputation & 0.0153 & 0.4442 & 0.7648 & 0.1975 & 0.970\\
 			Conditional Multiple Imputation & 0.0060 & 0.6779 & 0.6768 & 0.4631 & 0.959\\
 			Cox Based Multiple Imputation & 0.0061 &0.6537 & 0.6617 & 0.4274 & 0.966  \\\bottomrule				
 		\end{tabular}
 	\end{center}
 \end{table}
 
 \begin{table}
 	\caption{Case 4:  $X \sim Weibull(2,1/4)$ and censoring depends on  $Y$.}\label{tab:1Bd}
 	\begin{center}
 		\begin{tabular}{lcl*{5}{c}}
 %	\multicolumn{6}{c}{Independent Censoring} 
 	 \toprule
 			\textbf{Light Censoring} &  &  &  &  & \\ \midrule 
 				& &  & Simulation & & Coverage\\
 			$N=100$ & Bias & $SE(\hat{\beta})$ &  Error & MSE & Probability\\ \cmidrule(lr){1-6}
 			Actual data (No Censoring) & -0.0022 & 0.8335 & 0.8484 & 0.6947 & 0.949\\
 			Complete-case & 0.0157 & 1.0285 & 1.0365 & 1.0581 & 0.944\\
 			%Ad hoc Substitution  & -0.0551 & 0.8138 & 0.8151 & 0.6653 & 0.939\\
 			Mean Substitution & 0.0157 & 1.0298 & 1.0365 & 1.0607 & 0.945\\
 			Maximum Likelihood & 0.0109 & 0.8881 & 0.9122 & 0.7888 & 0.948\\
 			Conditional Single Imputation & 0.0045 & 0.8847 & 1.1293 & 0.7827 & 0.959\\
 			Conditional Multiple Imputation & 0.0030 & 1.0421 & 1.0471 & 1.0859 & 0.949\\ 
 			Cox Based Multiple Imputation & 0.0049 & 1.4071 &  1.4771 & 1.9800 & 0.966\\ \midrule\addlinespace
 			$N=500$ &  &  &  &  & \\  \cmidrule(lr){1-6}
 			Actual data (No Censoring) & 0.0012 & 0.3793 & 0.3843 & 0.1439 & 0.968\\
 			Complete-case & -0.0067 & 0.4684 & 0.4685 & 0.2194 & 0.947\\
 			%Ad hoc Substitution  & 0.0530 & 0.3682 & 0.3733 & 0.1383 & 0.948\\
 			Mean Substitution & -0.0067 & 0.4688 & 0.4685 & 0.2198 & 0.948\\
 			Maximum Likelihood & -0.0063 & 0.4070 & 0.4105 & 0.1657 & 0.953\\
 			Conditional Single Imputation & -0.0039 & 0.3992 & 0.7403 & 0.1594 & 0.966\\
 			Conditional  Multiple Imputation & 0.0025 & 0.4500 & 0.4796 & 0.2025 & 0.956\\ 
 			Cox Based Multiple Imputation & 0.0032 & 0.5062 & 0.5111 & 0.2562 &0.967\\\midrule\addlinespace
 			\textbf{Heavy Censoring} &  &  &  &  & \\ \midrule
 			$N=100$ &  &  &  &  & \\ \cmidrule(lr){1-6}
 			Actual data (No Censoring) & -0.0022 & 0.8335 & 0.8484 & 0.6947 & 0.949\\
 			Complete-case & 0.0156 & 1.4648 & 1.4558 & 2.4120 & 0.944\\
 			%Ad hoc Substitution $\sqrt2C$ & -0.0391 & 0.8675 & 0.9062 & 0.7541 & 0.939\\
 			Mean Substitution & 0.0156 & 1.4671 & 1.4558 & 2.1526 & 0.945\\
 			Maximum Likelihood & 0.0328 & 1.0092 & 1.0090 & 1.0196 & 0.944\\
 			Conditional Single Imputation & 0.0349 & 1.0006 & 1.2314 & 1.0024 & 0.933\\
 			Conditional Multiple Imputation & 0.0201 & 1.4001 & 1.4091 & 1.9607 & 0.966\\ 
 			Cox Based Multiple Imputation & 0.0450 & 1.0955 & 1.0998 & 1.2116 & 0.966 \\\midrule
 			$N=500$ &  &  &  &  & \\ \midrule
 			Actual data (No Censoring) & 0.0012 & 0.3793 & 0.3843 & 0.1439 & 0.968\\
 			Complete-case & 0.0102 & 0.6578 & 0.6611 & 0.4328 & 0.947\\
 			%Ad hoc Substitution $\sqrt2C$ & -0.0569 & 0.3947 & 0.3961 & 0.1590 & 0.952\\
 			Mean Substitution & 0.0102 & 0.6596 & 0.6611 & 0.4352 & 0.948\\
 			Maximum Likelihood & 0.0164 & 0.4672 & 0.4630 & 0.2185 & 0.952\\
 			Conditional Single Imputation & 0.0301 & 0.4436 & 0.6156 & 0.1977 & 0.966\\
 			Conditional Multiple Imputation & 0.0134 & 0.6867 & 0.6867 & 0.4717 & 0.966\\
 			Cox Based Multiple Imputation & 0.0141 & 0.6771 & 0.6846 & 0.4587 & 0.966\\ \bottomrule			
 		\end{tabular}
 	\end{center}
 \end{table}

\begin{figure}[]
   \centering
   \includegraphics[width=11cm]{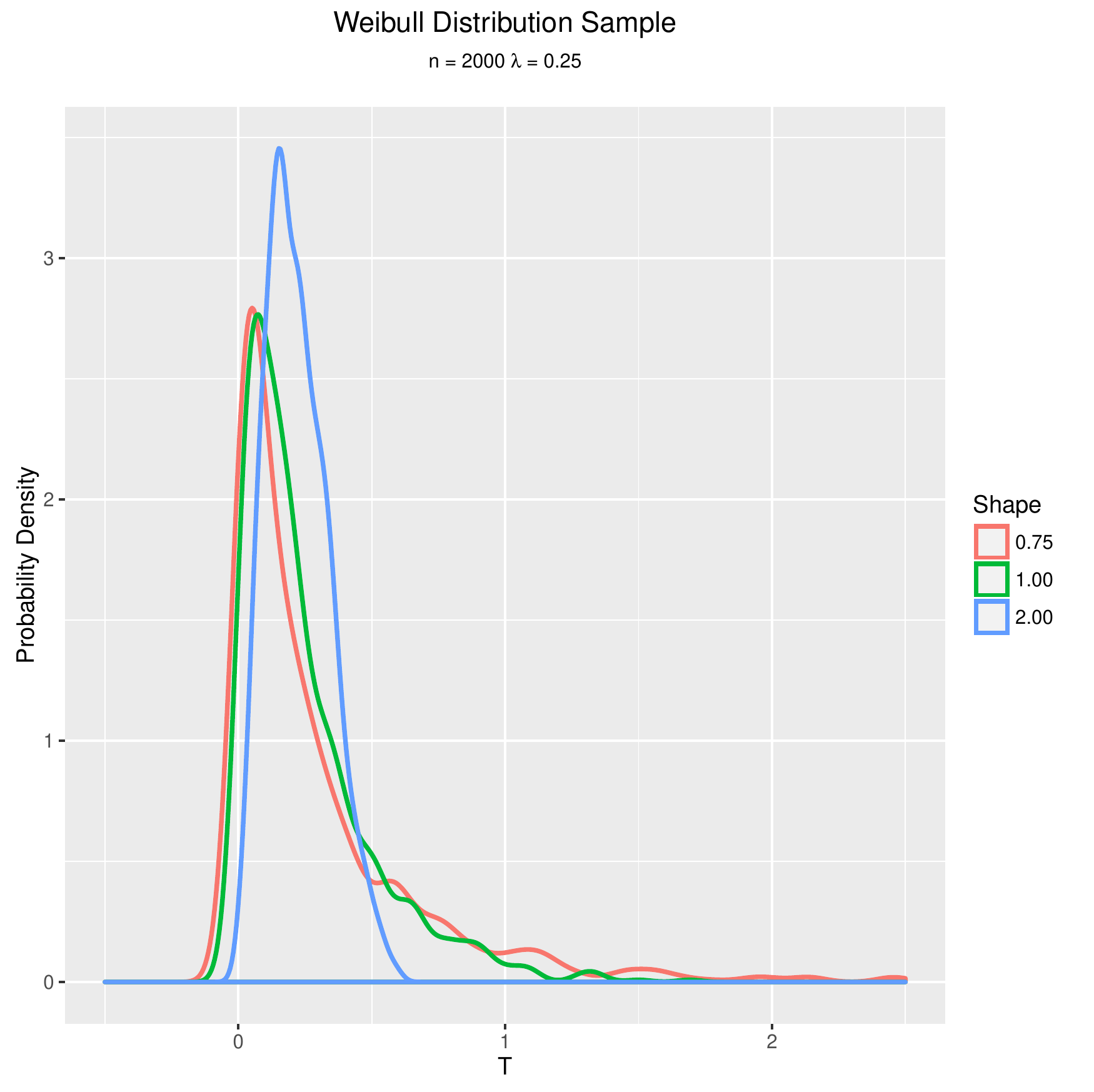}
    \caption{Distribution of the censored covariate as the function of the shape and scale parameters}\label{figure1}
     \end{figure}

 Tables \ref{tab:1Ba} and \ref{tab:1Bb} show that when the distribution of $X$ is highly skewed (see Figure \ref{figure1}), the parametric approach results in larger bias and MSE as compared to the conditional multiple imputation approach. Although the complete case is unbiased, deleting observations reduces the sample size, which results in an increased standard error and larger MSE as compared to both the maximum likelihood and the conditional multiple imputation methods.  Despite being unbiased, both the complete case and  the mean substitution methods are inefficient with  higher MSE as compared to the maximum likelihood approach ,conditional multiple imputation and the Cox multiple imputation approach.
 The single conditional imputation is unbiased and is more efficient than the mean imputation with smaller MSE because its underestimates the standard error when imputed values are used as true values with no uncertainty. All approaches resulted in acceptable coverage probabilities. 

Tables \ref{tab:1Bc} and \ref{tab:1Bd}  show that, when the distribution of $X$ is close to normal (see Figure \ref{figure1}), the maximum likelihood approach results in smaller bias and standard error. The log likelihood \eqref{eq:loglik} is derived under the normal distribution assumption. Therefore, the distribution of $X$ in Tables \ref{tab:1Bc}--\ref{tab:1Bd}, which is close to the true distribution from which the maximum likelihood method is based, provides a better and a more efficient parameter estimate than the  multiple imputation methods. The standard error and MSE is smaller than that of both the conditional multiple and Cox multiple imputation approaches. The other imputation methods are less efficient and more biased. Overall, it is worth mentioning that the Cox multiple imputation is more efficient than the conditional imputation when the data is well powered. As the sample size increases, this Cox multiple imputation is very efficient, which might be due to the fact that this approach uses one additional parameter  $Y$ in the imputation model as compared to the Kaplan-Meier based conditional imputation that does not involve the outcome in the imputation model.

%  {\color{blue} For the simulation section, I'm stopping to continue later. \underline{Date:} May 16 , 2016 at 6pm}
  
  \section{Illustrative example: Association between parent age of cardiac events and low density lipoprotein (LDL) in offspring.}\label{sec: apply}
 %According to the World Heart Federation and the World Health Organization, there about one billion people have high blood pressure (hypertension); of these, two thirds are in developing countries. Hypertension is one of the most important causes of premature death worldwide and the problem is growing. It is estimated that by 2025, 1.56 billion adults will be living with hypertension. Hypertension is the leading cause of cardiovascular disease worldwide and people with high blood pressure are more likely to develop complications {\color{red} shouldn't we say "from diabetes"? I am not sure, but "of diabetes" sounds off} of diabetes. Despite these studies little or nothing has been done to study the association between SBP and age of onset of cardiovascular event in parents.  
 
 According to the American Heart Association cardiovascular disease (CVD) is a multi-faceted disease that affects the heart or blood vessels. CVD includes hypertensive, rheumatic, congenital, and vulvar heart diseases as well as  cardiomyopathies, heart arrhythmias, carditis, aortic aneurysms, peripheral artery disease, venous thrombosis,  coronary death, myocardial infarction, coronary insufficiency, angina, ischemic stroke, hemorrhagic stroke, transient ischemic attack, peripheral artery disease, and heart failure.  It is the global leading cause of death, accounting for over 30\% of all deaths worldwide---approximately 17.3 million deaths per year. In the United States, someone dies from CVD every 39 seconds, with most of those deaths being attributed to coronary heart disease.\cite{schwalm2016resource, aha2012} 
 
 Though the death rate due to CVD has decreased slowly over the last 30 years, death from heart disease remains the leading cause of death in the United States, and caring for patients with poor cardiovascular health continues to be one of the largest burdens on the health care system today. From 1990 to 2009, CVD ranked first in the number of days for which patient received hospital care,\cite{cdc2010} yet 72\% of Americans do not consider themselves at risk for heart disease.\cite{aha2012}
 
 Associations have long been established between CVD and a wide variety of risk factors, including non-modifiable variable such as family history\cite{kannel1965comparison, keys1970coronary,neaton1992serum,manson1990prospective,donahue1987central,keil2000coronary,howard2000ldl}. Blood levels of low density lipoproteins  (LDL), one of the five major groups of lipoproteins categorized by density,  are regarded as a strong predictor of CVD.  
 To illustrate the methods proposed in this paper, we study the   association between LDL in offspring and age at onset of a clinically-diagnosed cardiovascular event in parents, using data from the Framingham Heart Study database and looking at both the Original and Offspring cohorts.\cite{kannel1979investigation}

The Framingham Heart Study (FHS) is an ongoing prospective study of the etiology of cardiovascular disease, among other prevalent diseases. The study began in 1948 and enrolled 5,209 participants (55\% women)  aged 28 and 62 years old residing in Framingham, Massachusetts as part of the original cohort who have been followed up to the present. In 1971, the Framingham Offspring Study was established with a sample of 5,124 men and women aged 5 to 70 years old who were either (genetic or adoptive) offspring or spouses of offspring of the original cohort\cite{dawber2015ii, mahmood2014framingham}. Study participants are examined routinely to update their health status information and potential risk factors. Standard clinical examinations included physician interview, physical examination, and laboratory tests, and continue to the present.  Participants in the original cohort have been followed biannually; there were 40 participants during the $32^{nd}$ Exam visit held in 2012--2014. In the offspring cohort, participants have been followed approximately every four years.  The Offspring Examination Cycle 9 covered the years 2011 to 2014 and had 2430 participants.

In this example, we performed two separate analyses, one for each parent, to evaluate the relationship between ages of CVD in parents and log(LDL) in offspring. Data gathered from the original cohort (Exam 12(1971--1974); 3,261 participants) and the offspring cohort(Exam 1(1971--1975);  5,124 participants) were used. We deleted all missing data and restricted the LDL to physician recorded values; this reduced the sample size to $n=2622$ (1,401 mothers and 1,221 fathers).  

Of the 1401 mothers in the final data set, 907 of them (i.e., 35.26\%) experienced a cardiovascular event  whereas  909 (i.e 74.45\%) out of 1,221 fathers experienced a cardiovascular event. The median age of CVD was 66 years and 63 years for mothers and fathers, respectively.  
 
 Results of the data analyses are provided in Tables \ref{data:apply1} and  \ref{data:apply2}. The results for the complete-case analysis, mean substitution, maximum likelihood, conditional single imputation and the conditional multiple imputation are consistent with the simulation results. With a larger sample, the assumption of normality for the censored covariate is met and the parametric method provides better estimates, along with smaller standard error. On the other hands, the results from the ad hoc substitution methods are inconsistent with the simulation results. This is because there is no scientific bases for such substitutions.  
 \section{Conclusion}
 %Censored data are common in many research areas including medicine, economics, and social sciences. An extensive literature has been developed for censored outcomes, especially in the studies of time-to-event outcomes where censoring may be due to loss of follow-up, drop out, or study termination.\cite{lagakos1979general, kleinbaum1996survival, klein2005survival} 
 Most of the literature on censored covariates deals with the issue of limits of detection, the point at which observations below this limit cannot be measured or detected and are instead recorded at the limit of detection value. \cite{bernhardt2014flexible,sattar2015frailty,rigobon2009bias} 
 In this paper, we considered the estimation of linear regression models when the covariate of interest is randomly censored. % Although the complete-case analysis method may yield valid inference, it can results in a substantial loss of both efficient and valuable information, especially if the proportion of censored observations is moderate or high. In addition, naive substitution methods which consist in replacing censored observations by a constant (mean, median, etc.) or  a function of highest observed measure $\tau$ (such as $\xi$, $\sqrt{2}\xi$, ${2}\xi$) is akin to fabricating data and should be discouraged.\cite{helsel2006fabricating,helsel2011statistics}
 We evaluated non parametric conditional imputation methods  based on the Kaplan-Meier estimate  to impute a censored covariate. We compared this non parametric approach based on Kaplan-Meier to  the regression from the full data (without censoring), the complete-case analysis, a naive {\it ad hoc} substitution (replacing censored values by the mean of the observed covariate values) and the maximum likelihood approach. 
 
 Parametric estimators were determined via maximum likelihood estimation method based on an underlying distribution assumption of the censored covariate. Throughout our simulations, we demonstrate that the naive {ad hoc} substitution method provides biased estimation of the regression parameter of the censored covariate.  As Helsel pointed out\cite{helsel2006fabricating}, these substitution methods are akin to fabricating data; they don't have any theoretical basis and should thus be discouraged. 
 
 The complete-case analysis method is  the widely used approach for handling censored predictors as it is easy to implement. The obvious pitfall of the complete-case approach is that it potentially sacrifices information by discarding observations. Although, this method leads to unbiased estimates under independent censoring, it can result in a substantial loss of power, especially under moderate to high percentages of censored observations. Under dependent censoring, complete-case analysis may lead to model misspecification due to selection bias if a group of subjects with similar characteristics do not experience the event of interest or leave a study before its  completion. 
 
The mean substitution approach is easy and looks reliable but a detailed analysis of this approach shows it has many short comings. We cannot always guarantee that the mean of the complete case will be greater than the time at censoring. One basic assumption of censored data is that, if the event is to occur,  it can only happen after the censored time. Furthermore, this approach does not make use of the available information, that is, the time at censoring.
 
 Using parametric methods requires prior knowledge or postulating a distribution model for the censored covariate. When the postulated parametric distribution of the censored covariate corresponds to the true distribution, the maximum likelihood estimation method and the nonparametric method via Kaplan-Meier estimation all provide consistent estimates, under independent censoring. Under dependent censoring, if the distribution of $X$ and $C$ are similar, these methods are efficient; however, if the distribution of $X$ and $C$ are dissimilar the MLE approach will be highly inefficient (as shown in section 2.1). Therefore, we propose the use of Kaplan-Meier nonparametric imputation in absence of prior knowledge of the distribution of censored covariate or when such a distribution cannot be accurately ascertained. On the other hand, if the sample size is large and the distribution of $X$ is a member of the exponential family, the MLE approach can be suitable.

\begin{table}[h]
	\caption{Relationship between Maternal age of onset of CVD and LDL in offspring}\label{data:apply1}
	\begin{center}
		%\begin{tabular}{p{4cm}lcccc}
		\begin{tabular}{lrccc}
		\toprule
			Method & Estimate & SE & P-value   \\ \midrule
			Complete-case (64.74\% of the data)
			& 0.0034 & 0.00012 & 0.0002  \\
			%Ad hoc Substitution & 0.0001 & 0.0005 &  0.9949 \\
			Mean Substitution & 0.0044 & 0.0012 &  0.0002 \\
			Maximum Likelihood & 0.1999 & 0.0797 & 0.0123\\
			Conditional Single Imputation & 0.0022 & 0.0009 &  0.0276 \\
			Conditional Multiple Imputation &  0.0023 & 0.0010 &   0.0284 \\  
			Cox Based Multiple Imputation &  0.0020 & 0.0009 & 0.0286\\\bottomrule
		\end{tabular}
	\end{center}
\end{table}

%\clearpage	
%\newpage	
%\vspace{12mm}
%\newpage
\begin{table}[h]
	\caption{Relationship between Paternal age of onset of CVD and LDL in offspring}\label{data:apply2}
	\begin{center}
		%\begin{tabular}{p{4cm}lcccc}
		\begin{tabular}{lrccc}
		\toprule
			Method & Estimate & SE & P-value  \\ \midrule
			Complete-case (74.45\% of the data)
			& 0.0024 & 0.0013 & 0.0675 \\
			%Ad hoc Substitution  & 0.0005 & 0.0006 & 0.3244 \\
			Mean Substitution & 0.0024 & 0.0013 &  0.0675 \\
			Maximum Likelihood & 0.7660 & 0.1480 &  $<.0001$ \\
			Conditional Single Imputation & 0.0018 & 0.0010 &  0.0742 \\
			Conditional Multiple Imputation &  0.0018 & 0.0011 &  0.0787 \\ 
			Cox Based Multiple Imputation & 0.0017 & 0.0010 & 0.0768\\\bottomrule
		\end{tabular}
	\end{center}
\end{table}

%\begin{multicols}{2}
\bibliographystyle{abbrv}
\bibliography{compare_epi}

%\end{multicols}
\pagebreak
\begin{appendices}
  \renewcommand\thetable{\thesection\arabic{table}}
  \renewcommand\thefigure{\thesection\arabic{figure}}     
  \section*{APPENDIX}  % use *-form to suppress numbering
 \setcounter{equation}{0}  % reset counter     
 \setcounter{section}{0} 
 \setcounter{table}{0} 
\numberwithin{equation}{section}
\numberwithin{table}{section}
%\noindent
\section{Likelihood function for a censored covariate}
Note that $\displaystyle Q=\sqrt{\sigma_{x}^{-2}+\beta_1^2\sigma^{-2}}=\sqrt{\frac{1}{\sigma_x^2} + \frac{\beta_1^2}{\sigma^2} }=\sqrt{\frac{\sigma^2+\sigma_x^2\beta_1^2}{\sigma^2\sigma_x^2} }$, which implies $\displaystyle \sigma^2\sigma_x^2=\frac{\sigma^2+\sigma_x^2\beta_1^2}{Q^2}.$

\allowdisplaybreaks \begin{eqnarray*}
f(Y_{i},X_{i}> c_{i})
&=&\int_{c_i}^{\infty} 
\exp\left\{ - \frac{\epsilon_i^2}{2\sigma^2} - \frac{1}{2} \log\left( 2\pi  \sigma^2 \right) \right\}
\exp\left\{ - \frac{(x_i-\mu_x)^2}{2\sigma_x^2} - \frac{1}{2} \log\left( 2\pi  \sigma_x^2 \right) \right\} \; \mathrm{d}x_i\\
% &=& \prod_{i=m+1}^n \left[1-\int_{-\infty}^{c_i - \mu_x} 
%\exp\left\{ - \frac{[y_i - \beta_0 - \beta_1(v_i+\mu_x)]^2}{2\sigma^2} - \frac{1}{2} \log\left( \sigma^2 \right) \right\}
%\exp\left\{ - \frac{v_i^2}{2\sigma_x^2} - \frac{1}{2} \log\left( \sigma_x^2 \right) \right\} \; \mathrm{d}v_i\right]
%\\
&=& \int_{c_i - \mu_x}^{\infty} 
\exp\left\{ - \frac{[y_i - \beta_0 - \beta_1(v_i+\mu_x)]^2}{2\sigma^2}  - \frac{v_i^2}{2\sigma_x^2}-\frac{1}{2} \log\left( 4\pi^2  \sigma^2 \sigma_x^2 \right)  \right\} \; \mathrm{d}v_i
\\
&=& \int_{c_i - \mu_x}^{\infty}
\exp\left\{ - \frac{\sigma_x^2[\epsilon_{i\mu_x} - \beta_1v_i]^2+\sigma^2v_i^2}{2\sigma^2\sigma_x^2} - \frac{1}{2} \log\left( 4\pi^2  \sigma^2 \sigma_x^2 \right) \right\} \; \mathrm{d}v_i
%\\
%&=&  \int_{c_i - \mu_x}^{\infty}
%\exp\left\{ - \frac{\sigma_x^2[(y_i - \beta_0 -\beta_1\mu_x)- \beta_1v_i]^2+\sigma^2v_i^2}{2\sigma^2\sigma_x^2} - \frac{1}{2} \log\left(4\pi^2   \sigma^2 \sigma_x^2 \right) \right\} \; \mathrm{d}v_i
\\
&=& \int_{c_i - \mu_x}^{\infty}
\exp\left\{ - \frac{\sigma_x^2[\epsilon_{i\mu_x}^2-2\epsilon_{i\mu_x} \beta_1v_i+\beta_1^2v_i^2]+\sigma^2v_i^2}{2\sigma^2\sigma_x^2} - \frac{1}{2} \log\left(4\pi^2   \sigma^2 \sigma_x^2 \right) \right\} \; \mathrm{d}v_i
\\
&=&  \int_{c_i - \mu_x}^{\infty}
\exp\left\{ - \frac{\epsilon_{i\mu_x}^2}{2\sigma^2}  - \frac{v_i^2 (\sigma^2+ \sigma_x^2\beta_1)-2 \sigma_x^2\epsilon_{i\mu_x} \beta_1v_i}{2\sigma^2\sigma_x^2} - \frac{1}{2} \log\left( 4\pi^2  \sigma^2 \sigma_x^2 \right) \right\} \; \mathrm{d}v_i
\\
&=&  \int_{c_i - \mu_x}^{\infty}
\exp\left\{ - \frac{\epsilon_{i\mu_x}^2}{2\sigma^2}  - \frac{1}{2}\left[v_i^2 Q^2-\frac{2 \epsilon_{i\mu_x} \beta_1v_i}{\sigma^2} \right]- \frac{1}{2} \log\left(4\pi^2   \sigma^2 \sigma_x^2 \right) \right\} \; \mathrm{d}v_i
\\
&=&  \int_{c_i - \mu_x}^{\infty}
\exp\left\{ - \frac{\epsilon_{i\mu_x}^2}{2\sigma^2}  - \frac{1}{2}Q^2\left[v_i^2 -2\frac{\epsilon_{i\mu_x} \beta_1}{\sigma^2Q^2} v_i\right]- \frac{1}{2} \log\left( 4\pi^2  \sigma^2 \sigma_x^2 \right) \right\} \; \mathrm{d}v_i
\\
&=& \int_{c_i - \mu_x}^{\infty}
\exp\left\{ - \frac{\epsilon_{i\mu_x}^2}{2\sigma^2} 
- \frac{1}{2} Q^2\left( v_i - \frac{\beta_1\varepsilon_{i\mu_x}}{\sigma^2 Q^2} \right)^2
+ \frac{\beta_1^2 \epsilon_{i\mu_x}^2}{2\sigma^4 Q^2}
- \frac{1}{2} \log\left( 4\pi^2  \sigma^2 \sigma_x^2 \right) \right\}
\; \mathrm{d}v_i
\\
&=&\exp\left\{ - \frac{\epsilon_{i\mu_x}^2}{2\sigma^2} 
+ \frac{\beta_1^2 \epsilon_{i\mu_x}^2}{2\sigma^4 Q^2}
- \frac{1}{2} \log\left( 4\pi^2  \sigma^2 \sigma_x^2 \right) \right\}
 \int_{c_i - \mu_x}^{\infty} 
\exp\left\{ - \frac{1}{2}Q^2\left( v_i - \frac{\beta_1\epsilon_{i\mu_x}}{\sigma^2 Q^2} \right)^2
\right\}
\; \mathrm{d}v_i
\\
&=&\exp\left\{ - \frac{\epsilon_{i\mu_x}^2}{2\sigma^2} 
+ \frac{\beta_1^2 \epsilon_{i\mu_x}^2}{2\sigma^4 Q^2}
- \frac{1}{2} \log\left( 2\pi  \sigma^2 \sigma_x^2 \right) \right\}
\left[ \frac{1}{Q}-\frac{1}{\sqrt{2\pi }}\int_{-\infty}^{c_i - \mu_x} 
\exp\left\{ - \frac{1}{2}Q^2\left( v_i - \frac{\beta_1\epsilon_{i\mu_x}}{\sigma^2 Q^2} \right)^2
\right\}
\; \mathrm{d}v_i\right]
\\
&=&
\exp\left\{ - \frac{\epsilon_{i\mu_x}^2}{2\sigma^2}
+ \frac{\beta_1^2 \epsilon_{i\mu_x}^2}{2\sigma^4 Q^2}
- \log\left( \sqrt{2\pi \sigma^2 \sigma_x^2 }\right) \right\}
\frac{1}{Q} \left[ 1-\Phi \left\{     
Q\left(  c_i - \mu_x - \frac{\beta_1\epsilon_{i\mu_x}}{\sigma^2 Q^2} \right)
\right\}\right]
\\
&=&\frac{1}{\sqrt{2\pi \sigma^2 \sigma_x^2 }}\frac{1}{Q}  \left[ 1-\Phi
\left\{
Q \left(
 c_i - \mu_x - \frac{\beta_1\epsilon_{i\mu_x}}{\sigma^2 Q^2}
\right) \right\}\right]
\exp\left\{ - \frac{\epsilon_{i\mu_x}^2}{2\sigma^2}
+ \frac{\beta_1^2 \epsilon_{i\mu_x}^2}{2\sigma^4 Q^2}
\right\}
\\
&=&
\frac{1}{\sqrt{2\pi (\sigma^2 + \beta_1^2 \sigma_x^2)}}
\left[ 1-\Phi
\left\{
Q \left(
 c_i - \mu_x - \frac{\beta_1\epsilon_{i\mu_x}}{\sigma^2 Q^2}
\right) \right\}\right]\exp\left\{ - \frac{\epsilon_{i\mu_x}^2}{2\sigma^2}
\left( 1 - \frac{\beta_1^2}{\sigma^2 Q^2} \right) \right\}
\\
&=&
\frac{1}{\sqrt{2\pi (\sigma^2 + \beta_1^2 \sigma_x^2)}}
\Phi
\left\{
Q \left(
 \mu_x + \frac{\beta_1\epsilon_{i\mu_x}}{\sigma^2 Q^2}-c_i 
\right) \right\}\exp\left\{ - \frac{\epsilon_{i\mu_x}^2}{2\sigma^2}
\left( 1 - \frac{\beta_1^2}{\sigma^2 Q^2} \right) \right\}
\end{eqnarray*}
The log likelihood equation \eqref{eq:loglik} from the main text becomes
\allowdisplaybreaks \begin{eqnarray}\label{eq:loglikf}
\log(L) &=&g(\delta)-\frac{\delta}{2}\sum\limits_{i=1}^{m}\bigg({\log(\sigma^2\sigma_x^2)+\frac{\epsilon_{i}^2}{\sigma^2}+\frac{(x_{i}-\mu_x)^2}{\sigma_{x}^2}}\bigg)-\frac{(1-\delta)}{2}\sum\limits_{i=m+1}^{n}\left\{ \log(\sigma^2 + \beta_1^2 \sigma_x^2)\right.\nonumber\\
&&+ \left.\frac{\epsilon_{i\mu_x}^2}{\sigma^2}
\left( 1 - \frac{\beta_1^2}{\sigma^2 Q^2} \right) -2\log\left[\Phi
\left\{
Q \left(
 \mu_x + \frac{\beta_1\epsilon_{i\mu_x}}{\sigma^2 Q^2}
-c_i\right) \right\}\right] \right\}\nonumber\\
&=&g(\delta)-\frac{m}{2}\delta\log(\sigma^2\sigma_x^2)- \frac{n-m}{2}(1-\delta)\log(\sigma^2 + \beta_1^2 \sigma_x^2) -\frac{\delta}{2} \sum\limits_{i=1}^{m}\bigg({\frac{\epsilon_{i}^2}{\sigma^2}+\frac{(x_{i}-\mu_x)^2}{\sigma_{x}^2}}\bigg)\nonumber\\
&&-\frac{(1-\delta)}{2}\sum\limits_{i=m+1}^{n}\left\{ \frac{\epsilon_{i\mu_x}^2}{\sigma^2}
\left( 1 - \frac{\beta_1^2}{\sigma^2 Q^2} \right) -2\log\left[\Phi
\left\{
Q \left(
 \mu_x + \frac{\beta_1\epsilon_{i\mu_x}}{\sigma^2 Q^2}
-c_i\right) \right\}\right] \right\}
\end{eqnarray}
where the constant term $g(\delta)=-\frac{1}{2}[2m\delta+(n-m)(1-\delta)]\log(2\pi).$
%{\color{blue} I have stopped here. Date: April 7 , 2016 at 1:49am}

% \section*{Acknowledgements}
% 
% \section{Appendix}
\pagebreak
 \section{SAS Code: Parametric model}
 \begin{SAScode*}
 proc NLMIXED data=data-set;
 parms mux=  sigmax= sigma=  alpha=  beta= ;
 Q=sqrt(1/sigmax**2+beta**2/sigma**2);
 e=y-alpha-beta*time; emu=y-alpha-beta*mux;
 if censored=0 then 
   LL=(-e**2/sigma**2/2-log(sigma**2*sigmax**2)/2
                        -(time-mux)**2/sigmax**2/2);
 if censored=1 then 
   LL=log(sqrt(sigma**2+beta**2*sigmax**2)**-1*probnorm(Q*(-time+mux
   +sigma**-2*Q**-2*beta*emu))*exp(2**-1*sigma**-4*Q**-2*beta**2*emu**2 -2**-1*sigma**-2*emu**2));
 
 model y~general(LL);
 run;
 
 	\end{SAScode*}
% \section{SAS Code: Non-Parametric model}	
\end{appendices}
%\begin{acks}
\section*{Acknowledgement}
The data for this study was approved by The University of Texas Health Science Center Institutional Review Board and was made available with the help of the Biologic Specimen and Data
 Repository Information Coordinating Center.  Request for access to the Framingham Pedigree was approved by the Framingham Executive Committee.
 The authors acknowledge research support from the National Institutes of Health (NIH). \\
 R.A. Matsouaka was supported by the National Center for Advancing Translational Sciences of the National Institutes of Health under Award Number UL1TR001117. \\The content of this paper is solely the responsibility of authors and does not necessarily represent the official view of the National Institutes of Health. \\\\
 {\it Conflict of Interest}: None declared.	
 
%The authors thank the editor, the associate editor, and the reviewers for their comments and suggestions, which have helped tremendously to improve the quality of this manuscript.
%\end{acks}
\end{document}